\documentclass[useAMS,usenatbib]{mn2e}

\title[AGN dust tori at low and high luminosities]{AGN dust tori at low and high luminosities}
\author[S. F. H\"onig and T. Beckert]{S. F. H\"onig\thanks{E-mail:
shoenig@mpifr-bonn.mpg.de} and T. Beckert\\
Max-Planck-Institut f\"ur Radioastronomie, Auf dem H\"ugel 69, 53121 Bonn, Germany}

\begin{document}

\newcommand{\Lestd}{L_{\rm edd}^{(\rm std)}}
\newcommand{\Lehom}{L_{\rm edd}^{(\rm smooth)}}
\newcommand{\Lecl}{L_{\rm edd}^{(\rm cl)}}
\newcommand{\Ledd}{L_{\rm edd}}
\newcommand{\ergs}{\rm erg\,s^{-1}}

\newcommand{\LLedd}{\ell_{\rm Edd}}

\newcommand{\Msev}{M_7}
\newcommand{\Mbh}{M_{\rm BH}}

\newcommand{\Ms}{\rm M_\odot}
\newcommand{\Msy}{\rm M_\odot\,yr^{-1}}
\newcommand{\MdAD}{\dot{M}_{\rm AD}}
\newcommand{\Mdtor}{\dot{M}_{\rm Torus}}
\newcommand{\Mdout}{\dot{M}_{\rm outflow}}

\newcommand{\csound}{c_{\rm s}}
\newcommand{\rsub}{r_{\rm sub}}
\newcommand{\ro}{r_{\rm out}}
\newcommand{\rpc}{r_{\rm pc}}
\newcommand{\rmax}{r_{\rm max}}

\newcommand{\Rcl}{R_{\rm cl}}
\newcommand{\Rclmax}{R_{\rm cl,max}}
\newcommand{\Mcl}{M_{\rm cl}}
\newcommand{\pv}{\Phi_{\rm V}}
\newcommand{\N}{\mathcal{N}}
\newcommand{\Nc}{N_{\rm c}}
\newcommand{\tcl}{\tau_{\rm cl}}
\newcommand{\nH}{n_{\rm H}}

\date{Accepted 2007 June 26. Received 2007 June 06; in original form 2007 February 08}

\pagerange{\pageref{firstpage}--\pageref{lastpage}} \pubyear{2007}

\maketitle

\label{firstpage}

\begin{abstract}
A cornerstone of AGN unification schemes is the presence of an optically
and geometrically thick dust torus. It provides the obscuration to explain
the difference between type 1 and type 2 AGN.
We investigate the influence of the dust distribution on the Eddington
limit of the torus. For smooth dust distributions, the Eddingtion
limit on the dust alone is 5 orders of magnitudes below the limit for
electron scattering in a fully ionized plasma, while a clumpy dust torus
has an Eddington limit slightly larger than the classical one.
We study the behaviour of a clumpy torus at low and high AGN luminosities.
For low luminosities of the order of $\sim 10^{42}\,\ergs$, the torus changes its
characteristics and obscuration becomes insufficient. In the high luminosity regime,
the clumpy torus can show a behaviour which is consistent with the
``receding torus'' picture. The derived luminosity-dependent fraction
of type-2-objects agrees with recent observational results. Moreover,
the luminosity-dependent covering factor in a clumpy torus may explain
the presence of broad-line AGN with high column densities in X-rays.
\end{abstract}


\section{Introduction}

The widely accepted unification scheme for active galactic nuclei (AGN) proposes that the central accretion disk and broad line region (BLR) are
surrounded by a geometrically thick dusty torus \citep[e.g.,][]{Anton93}. The dust in the torus obscures the accretion disk and BLR for lines of sight which pass
through the torus, while they are visible otherwise. The spectral energy distributions (SED) of most Quasars and AGN in Seyfert galaxies have a
pronounced secondary peak in the mid-infrared \citep[e.g.,][]{San89,Elv94}, which is interpreted as thermal emission by hot dust in the torus. The dust is heated by the primary optical/UV continuum radiation, and the torus extends from the dust sublimation radius outwards \citep{Bar87}. To date, the thermal dust emission from the torus has only been spatially resolved by infrared interferometric techniques for the torus in NGC~1068 \citep{Wei04,Jaf04} and Circinus \citep{Tr06}. The actual geometry and physical properties (e.g., dynamics and dust composition) are, thus, still unknown. In particular, the geometrical thickness, which determines the covering (obscuration) factor, remains a puzzle: Although axisymmetric, rotating gas and dust configurations with cooling will form thin disks, the torus should keep an aspect ratio $H/r \ge 0.5$ for most of the AGN activity phase.

Besides the geometrical thickness, the dynamics of dust in the torus is strongly affected by radiation pressure from the primary AGN luminosity.
It has been suggested that the (radiation) pressure by starformation inside the torus may solve the problem of its geometrical thickness
\citep[e.g.,][]{Ohsu99, Wad02}. In a competing scenario, the torus consists of a large number of small, self-gravitating, dusty molecular
clouds which form a clumpy torus \citep{Kro88, Bec04}.

In this article, we descibe the consequences of AGN radiation pressure for the clumpy torus model. In Sect.~2, we describe the Eddington limit on gas and dust.
In Sect.~3 and 4, we introduce a lower and an upper luminosity limit for the torus, respectively. We summarize our results in Sect.~5.

\section{The Eddington limit for the torus} \label{EddLimTor}

In the classical picture, the Eddington limit is defined as the state when gravity of the enclosed mass balances the radiation pressure from
the central source, so that
\begin{equation}\label{Ledd_classic}
\Ledd = 4 \pi c G M(r)\cdot\frac{m}{\sigma}
\end{equation}
Here $G$ is the gravity constant, $c$ is the speed of light, and $m$ and $\sigma$ are the mass and the cross section of the particle which is exposed to the radiation.
For a fully ionized plasma arround a black hole, the inverse opacity $m/\sigma = \kappa^{-1}$ is dominated by the proton
mass and Thomson scattering of electrons. This changes in the region of the torus where dust is mixed with gas.
For reference, we use $\kappa_0=\sigma_T/m_p=0.4\,{\rm cm^2/g}$ for the fully ionized gas.
Assuming that gravity is dominated by the black hole mass $M=\Mbh = \Msev\times10^7 M_\odot$,
we obtain the classical Eddington limit
\begin{equation}\label{Ledd_std}
\Lestd=1.26\times10^{45}\,\ergs\cdot\Msev
\end{equation}

The time-averaged AGN luminosity is expected to scale with the mass accretion rate in the accretion disk by $L=\eta \MdAD c^2$.
In a stationary scenario, the accretion disk itself is fueled by mass transported through the torus from galactic scales. The mass transport rate through the torus
$\Mdtor$ is related to $\MdAD$ via $\MdAD=\Mdtor-\Mdout$. This relation considers mass loss in an outflow or jet during the accretion process from the
torus towards the inner accretion disk. With $\tau=1-\Mdout/\Mdtor$, we obtain
\begin{equation}\label{LMdot}
L=\eta\tau \Mdtor c^2.
\end{equation}
The combination of theory and observation for radiative efficient accretion with outflows suggests $0.01\la\eta\la0.1$ and $0.1\la\tau\la1$ \citep[e.g.,][]{Emm92,Pel92}. In the following
we will use a representative value of $\eta\tau=0.05$. From Eqn.~(\ref{LMdot}), we obtain $\Mdtor=0.4\,\Msy\times L_{45}$. Here $L_{45}$ is the bolometric luminosity in units of $10^{45}\,\ergs$.

We will now investigate the accretion properties for a dusty medium in the AGN torus considering a smooth dust distribution and
a clumpy structure, respectively.

\subsection{Smooth dust distribution}\label{sdd}

Radiative transfer simulations of AGN dust tori frequently use dust which is smoothly distributed \citep[e.g.,][]{Pie92,Gra94,Sch05}. The radiation
which comes from the central AGN directly acts on the dust grains in the torus. The absorption cross section of the dust grains in the optical and UV regime
can be approximated by their geometrical cross section $\sigma=\pi r_{\rm Dust}^2$. Standard size distributions assume dust grain sizes between 0.025\,\micron\,
and 0.25\,\micron\, \citep[e.g.,][]{Mat77}. Using a typical dust grain density of $2-3\,{\rm g/cm^3}$, we obtain an opacity
$\kappa_{\rm Dust} \sim 3-40\times10^4\,\kappa_0$. Using $\kappa_{\rm Dust}$ in Eqn.~(\ref{Ledd_classic}), the Eddington luminosity for smoothly
distributed dust becomes
\begin{equation}\label{lehom}
\Lehom = 0.3-5\times10^{40}\,\ergs\cdot\Msev
\end{equation}
Due to the large value of $\kappa$, the Eddingtion luminosity decreases by 5 orders of magnitude. As a consequence, typical AGN luminosities of $10^{45}\,\ergs$ would
require a black hole mass of $10^{12}\,M_\odot$. This is, however, inconsistent with observed
$M_{\rm BH}/L_{\rm bol}$-ratios \citep[e.g.,][]{Kas00,Woo02}. As a consequence the dust can not be gravitationally bound to the black hole.

The result with the approximated $\kappa_{\rm Dust}$ is consistent with recent opacity calculations for a more realistic dust and gas mixture \citep[e.g.,][]{Sem03}. They show that dust opacities for UV temperatures, which are dominating the AGN accretion disk radiation, are about 4 orders of magnitude larger than the Thomson opacity $\kappa_0$. In principle, IR photons coming from more external regions of the torus act as a counter force to the the UV photon pressure from the AD. However, the IR opacity of the gas and dust mixture is $\sim10^3$ times smaller than the UV opacity, resulting in an insignificant effect of IR photons when compared to the dominating UV photons. Furthermore, the geometry of the torus causes the diffuse IR torus radiation to act more or less isotropically on the inner wall of the torus (at least in the torus plane), so that a possible IR counter pressure is even weakened. Contrary, it is rather expected that the main effect of the IR photon pressure inside the torus is a vertical thickening \citep{Kro07}.

Eqn.~(\ref{lehom}) considers dust grains which are decoupled from the gas. However, even for a perfect coupling of gas and dust in the torus (no drift of the dust relative to the gas), the usually assumed mass ratio of gas to dust of 100 raises the limit $\Lehom$ to only $0.001$ of the classical limit in Eqn.~(\ref{Ledd_std}). The limit $\Lehom$ is valid for an optically thin gas dust mixture, while AGN tori are necessarily optically thick, providing self-shilding of most of the torus against the AGN radiation. This creates a radiation pressure gradient at the inner boundary layer (width $\tau_{\rm UV} \sim 1$) of the torus. The corresponding outward force on this layer is $L/\Lehom \sim \Lestd/\Lehom \sim 10^3$ times stronger than the gravitational pull of the central black hole. This outward force would have to be counterbalanced by an enormous inward pressure gradient in the inner boundary layer.

\subsection{Clumpy dust torus}\label{clumpytorus}

An alternative model to smoothly distributed dust was proposed by \citet{Kro88}. They argue that most of the gas and dust in the torus around an AGN has
to be arranged in optically thick, self-gravitating clouds. \citet{Voll04} and \citet{Bec04} presented a stationary accretion model for the clumpy torus (hereafter:
SA model), including relations for torus
and dust cloud properties. The main idea behind this model is that clouds are very compact with a large optical depth in the UV due to dust which provides self-shielding against the AGN radiation and allows the clouds interior to be cold. Dust
grains on the directly illuminated sides of the clouds are exposed to the AGN and individual dust grains are potentially accelerated and expeled from the cloud due to the radiation pressure. Both magnetic fields and dynamical friction of grains in the gas phase of the cloud \citep[see, e.g.,][Sec. 9]{Spitzer78} can prevent this and transfer the momentum to the gas. In a self-gravitating cloud the radiation pressure is therefore received by the whole cloud. Thus, the torus is limited by the radiation pressure from the central AGN acting on the dust clouds instead of single grains,
so that we can define a cloud opacity $\kappa_{\rm cl}=\pi \Rcl^2/\Mcl$.

The SA model assumes the clouds to be self-gravitating, so that the free-fall time equals the sound crossing time $R_{\rm cl}/\csound$. This provides a linear relation between cloud mass $\Mcl$ and $R_{\rm cl}$. These clouds should be stable against tidal forces in the gravitational field of the central black hole. This requires $R^3_{\rm cl}/r^3 \le 2 \Mbh/\Mcl$. When combining both limits \citep[see][]{Bec04}, one finds an upper limit for the cloud size
\begin{equation} \label{cloudradius}
\Rclmax = \frac{\pi}{\sqrt{8G}}\cdot\frac{\csound r^{3/2}}{M^{1/2}}
\end{equation}
and a corresponding mass
\begin{equation}\label{cloudmass}
\Mcl=\frac{\pi^2\csound^2}{8G}\Rcl \;.
\end{equation}
Here $c_{\rm s}$ is the cloud-internal speed of
pressure waves which is of the order of $1\,{\rm km\,s^{-1}}$. We use this value as the unit for $\csound$ in the following. This speed characterizes
the cloud internal pressure which is required to balance self-gravity, and can be understood as
the speed of supersonic turbulence in the clouds. Alternatively, the clouds may be magnetically supported \footnote{The mass density of clouds from Eqn. (\ref{cloudradius}) and (\ref{cloudmass}) is $\rho = 1.6\times10^{-16}$\,g cm$^{-3}\,M_7 r_{\rm pc}^{-3}$. A dynamically relevant $B$-field will have a strength of a few mG. This leads to gyration times of grains shorter than the dynamical time of the clouds $\Rclmax/c_s$. The large densities $n_H \sim 8\times10^7$\,cm$^{-3}\,M_7 r_{\rm pc}^{-3}$ are sufficient to effectively transfer momentum from grains to the gas and to limit the drift velocity of grains to about $c_s$. Both mechanisms support the above claim that radiation pressure acts on the whole cloud.}.
Due to their large cross section, these clouds dominante the absorption, scattering, and IR re-emission.
From the relations for $\Rcl$ and $\Mcl$, we get an upper envelope for the opacity of
$\kappa_{\rm cl}= 0.7\kappa_0 \cdot \rpc^{3/2}/(\csound \Msev^{1/2})$. The distance from the black hole $\rpc$ is measured in pc and the speed
$c_{\rm s}$ in km s$^{-1}$.  For clouds smaller than the shear limit, the opacity $\kappa_{\rm cl}\propto \Rcl$ becomes
smaller. With Eqn.~(\ref{Ledd_classic}), we obtain the Eddington limit for clouds in a clumpy torus, which are directly
exposed to the primary AGN radiation,
\begin{equation}\label{Ledd_clumpy}
\Lecl=1.78\times10^{45}\,\ergs\cdot \frac{\csound \Msev^{3/2}}{\rpc^{3/2}}\;.
\end{equation}
This is of the same order as in the classical Eddington limit (Eqn.~\ref{Ledd_std}) and is consistent with observed AGN luminosities and black hole masses.
Since $\Ledd \propto \kappa^{-1}$, the Eddington limit for small clouds is even larger than in Eqn.~(\ref{Ledd_clumpy}).

Eqn.~(\ref{Ledd_clumpy}) shows that $\Lecl \propto r^{-3/2}$. This implies that at larger distances, self-gravitating clouds which
are directly exposed to the AGN radiation become unbound by the radiation pressure. Thus, distant clouds have to be shielded against the AGN radiation by clouds at small radii. As a consequence,
there should be no significant vertical flaring for a clumpy torus; i.e., we expect $H/r\approx{\rm const}$.
Further consequences of this behaviour will be discussed in Sect.~\ref{Lhigh}.

\section{The torus at low AGN luminosities}\label{S_Llow}

In the previous section, we argued that the Eddington limit for clumpy dust tori is well in agreement with the range of observed AGN luminosities and black
hole masses. To be clumpy, a torus requires a small volume filling factor $\pv\ll1$ for the dusty clouds. In the context of a SA model, \citet{Bec04} find a mass transport rate through the torus $\Mdtor = 3 \pi \nu \Sigma$, where $\nu= \frac{\tau}{1+\tau^2} H^2 \Omega_{\rm Kepler}$ is the effective viscosity for a torus and $\Sigma$ is the surface density. For an obscuring torus, the scale height, $H$, cannot be smaller than the mean free path of clouds $H \ge l = (4/3)R_{\rm cl}/\pv$. Otherwise the torus would become transparent for AGN photons. The parameter $\tau$ in the viscosity prescription measures the ratio $\tau=l/H$. The geometric thickness of the torus and the viscosity is maximised for $\tau = 1$. For $H \gg l$ the cloud density in the torus growth rapidly and the torus would collapse to a thin disk. We therefore adopt $H=l$ for a working model. After replacing the mean free path by the appropriate expression from \citet{Bec04}, we get $\pv$ in terms of $\Mdtor$,
\begin{equation}\label{PhiV}
\pv=\frac{\pi^{7/4}}{\sqrt{6G}}\cdot\frac{\csound^{3/2}}{\Mdtor^{1/2}}\,.
\end{equation}
The volume filling factor only depends on the mass transport rate through the torus, which we parametrized by
$\Mdtor=0.4\,\Msy\times L_{45}\cdot(\eta\tau/0.05)^{-1}$ (see Sect.~\ref{EddLimTor}). By substituting $\Mdtor$, we obtain a hard lower luminosity for the existence of an obscuring torus according to the SA model,
\begin{equation}\label{Llow}
L_{\rm low} = 5\times10^{42}\,\ergs \cdot\left[\frac{\eta\tau}{0.05}\right] \,.
\end{equation}
at which $\pv=1$. For clumpy obscuring tori as described, it is necessary that the AGN luminosity is $L\gg L_{\rm low}$. If $L \ga L_{\rm low}$, the volume filling factor becomes $\pv\rightarrow1$. At this point, the SA model would require that the torus collapses to a geometrically thin disk. As a consequence, most of the dust would be driven away (see Sect.~\ref{sdd}). It is, however, known that this situation can be avoided: Lower luminosities go along with lower accretion rates. \citet{Voll04} showed that for low mass accretion rate, a clumpy and almost transparent ($l \gg H$) circumnuclear disk (CND) can form similar to what has been found around the central black hole in our Galaxy \citep{Gue87}. The difference between the clumpy torus and the CND is that the latter one loses most of its obscuration properties while there can still be IR reprocessing.

Several observational studies show that at about $10^{42}\,\ergs$, the $L_{\rm bol}-L_{\rm MIR}$- or $L_{\rm X}-L_{\rm MIR}$-relation
show a significant change in behaviour compared to higher luminosities \citep[e.g.,][]{Lut04,Hor06}. Apparently, the main source of MIR emission
at $L\sim10^{42}\,\ergs$ is not the proposed, geometrically thick torus anymore.

A similar low-luminosity limit has been found for models where the dust clouds are not produced in a torus but released into a wind
from an accretion disk \citep{Eli06}. The cutoff at lower luminosities is a result of the fact that the mass outflow rate in the wind cannot exceed the mass accretion rate in the disk. Taking the same $\tau$ and $\eta$ as used in \citet{Eli06}, we obtain $L_{\rm low}=2\times10^{42}\,\ergs$.

\section{The dust torus in the high luminosity regime}\label{Lhigh}

In Sect.~\ref{clumpytorus}, we have shown that the Eddington luminosity for clouds in the clumpy torus, $\Lecl$, depends on the cloud-AGN distance as $r^{-3/2}$. A large fraction of AGN radiate close to or at the classical Eddington limit for Thomson scattering \citep{Mac01}.
We therefore scale the actual luminosity to the classical limit $L =\Lestd\LLedd$, where $\LLedd\le 1$ is the Eddington ratio for the AGN.
Once $\Lecl$ becomes smaller than $L$, the clouds of corresponding $\Rcl(r)$ which are directly exposed to the radiation of the central source can no longer resist the radiation pressure. This defines the condition $L/\Lecl<1$ for the existence of dust clouds of radius $\Rcl(r)$ in the AGN radiation field. Since $\Lecl$
is $r$-dependent, we obtain a maximum distance from the AGN, $\rmax(\Rcl)$, at which the largest clouds can withstand the radiation pressure:
\begin{equation}
\rmax(\Rcl)=1.3\,{\rm pc} \cdot \LLedd^{-2/3} \csound^{2/3}\Msev^{1/3}\cdot\left(\frac{\Rcl}{\Rclmax}\right)^{-2/3}
\end{equation}
The factor $(\Rcl/\Rclmax)\le1$ accounts for different cloud radii $\Rcl$ up to the sheer limit $\Rclmax$ (see Eqn.~(\ref{cloudradius})). As mentioned in
Sect.~\ref{clumpytorus}, clouds at distances $r>\rmax$ need to be shieled, so that no vertical flaring should occur beyond $\rmax$. We want to note that
$\rmax$ does {\it not} refer to an outer radius of the torus. It relates the maximum possible size of a dust cloud to the radiation pressure from the central
AGN.

A proper scaling  of the limiting radius, $\rmax$, is in units of the dust sublimation radius which presumably sets the inner radius of the torus. The sublimation radius is depending on the actual dust chemistry and grain sizes. A well-referenced estimation of the $\rsub$ was introduced by \citet{Bar87}, providing
$\rsub = 0.4\,{\rm pc}\times L_{45}^{1/2} T_{\rm sub;1500}^{-2.8} a_{0.05}^{-1/2}$, where sublimation of graphite grains of radius $0.05\,{\rm\mu m}$ and a sublimation temperature of 1500\,K is assumed. For different grain sizes and chemistry, the sublimation temperature and the temperature exponent change. Thus, in the case of silicate grains with similar grain size, the sublimation radius is larger by about a factor of about 3. While the actual chemistry, grain sizes and sublimation temperatures around AGN are still a matter of debate \citep[e.g.,][]{Bar92,Sit93,Kis07}, reverberation measurements of type 1 AGN support $\rsub \propto L^{1/2}$ \citep{Sug06}. In the following, we will use the simplyfied relation $\rsub = 0.5\,{\rm pc}\times L_{45}^{1/2}$, keeping in mind the uncertainty due to the actual dust mixture.

Finally, this allows for a comparison of $\rmax$ with the inner boundary of the torus:
\begin{equation} \label{maxrad}
\frac{\rmax}{\rsub}=2.3\cdot\csound^{2/3}L_{45}^{-1/6}\cdot\LLedd^{-1}\left(\frac{\Rcl}{\Rclmax}\right)^{-2/3}
\end{equation}
Interestingly, $\rmax/\rsub$ is approximately of the order of unity. That means that at $L\ga L_{45}$, the maximum dust cloud size can be limited by
the radiation pressure rather than the shear limit. Which mechanism dominates for an individual AGN depends on its actual $L/M$. We will, thus, distinguish between
radiation-limited tori (AGN with higher $\ell$ or $L$) and shear-limited tori (AGN with lower $\ell$ or $L$) in the following.

An effect of Eqn.~(\ref{maxrad}) is a change in obscuration properties with higher luminosities. In Sect.~\ref{S_Llow}, we briefly summarized the results from the SA model which requires $H=l=4/3\,\Rcl/\pv$. Contrary to the lower AGN luminosities, $\Rcl$ is now defined by the radiation limit (Eqn.~(\ref{maxrad}))
instead of the shear limit (Eqn.~(\ref{cloudradius})). From $\Rcl \propto L^{-1/4}$ (Eqn.~(\ref{maxrad})) and $\pv \propto \Mdtor^{-1/2} \propto L^{-1/2}$
(Eqn.~(\ref{LMdot})), we obtain $H \propto L^{1/4}$. Since the torus is expected to have $H/R={\rm const}$, the thickness of the torus is determined at the
reference distance $R=\rsub\propto L^{1/2}$. This results in an average thickness of the torus,
\begin{equation}\label{HR_highL}
H/R \propto L^{-1/4}.
\end{equation}
Thus, we expect to see more type 1 AGN at higher luminosities for objects which have radiation-limited clumpy tori.

This result can be interpreted in the framework of the ``receding torus'' \citep{Law91}: It has been observed that for high-luminosity sources, the
observed hydrogen column density is lower than in low-luminosity sources \citep[e.g.,][]{Ued03,Bar05,LaF05,Aky06}. This is interpreted as a
decrease of the covering factor of the torus with luminosity; i.e., $H/R$  appears to be anti-correlated with $L$. Recently, \citet{Sim05} analysed the
luminosity-dependence of the type 1 and type 2 AGN fraction by combining the results from different surveys. While the original
receding torus \citep{Law91} predicts a type-2-fraction $f_2\propto L^{-1/2}$, a better fit was found for a situation where the height of the torus
depends on luminosity. From Eqn.~(\ref{HR_highL}), we would approximately expect $f_2\propto L^{-0.25}$ for radiation-limited dust tori. This is remarkably
close to the correlation $f_2\propto L^{-0.27}$ derived by Simpson. We note, however, that for objects with low $\LLedd$, the $L^{-1/4}$-dependence should not hold but can even be inverted (see Eqn.~{cloudradius}). This would imply that it depends on the actual AGN sample properties if a
receding torus is observed or not.

Several authors noted that the UV-to-IR dust extinction in AGN is lower than what would be inferred from the X-ray column density, in particular referring to
AGNs showing broad-lines in the optical and significant absorption in the X-rays \citep[e.g.,][]{Wil02,Per04,Bar05}. In the clumpy torus model,
the AGN's X-ray emission region is completely obscured by an average number, $N$, of optically thick clouds along a line of sight passing through the torus. Since the size of the X-ray
source is smaller than the typical cloud radius, $R_{\rm X}<\Rcl$, the X-ray column density is $n_{\rm X}\sim N\cdot n_{\rm cl}$, where $n_{\rm cl}$
denotes the column density of an individual cloud. On the other hand, the BLR is only fractionally (and statistically) obscured by a number of $N$ optically
thick clouds, since $R_{\rm BLR} > \Rcl$. As a result, the average optical depth in the optical wavelength range, $\tau_{\rm V}$, can
be approximated by $\tau_{\rm V}\sim N$ \citep{Nat84}. Thus, the inferred optical depth from the X-rays, $ N\cdot\tau_{\rm cl}$, overestimates the measured optical depth $\tau_{\rm V}\approx N$ for a more extended emission region like the BLR.
Here, $\tau_{\rm cl}\gg1$ is the optical depth of an individual dust cloud.
Furthermore, if $N$ is sufficiently small, there exists a high probability that parts of the BLR are directly visible to the observer, while the
X-ray source is still obscured ($R_{\rm BLR} > \Rcl > R_{\rm X}$). This effect should become stronger at higher AGN luminosities when the clumpy tori become
radiation-dominated, since $N\propto l^{-1}\propto L^{-1/4}$. Actually, \citet{Per04} report that $\sim10\%$ of their observed broad-line AGN show
high column densities, all of them having X-ray luminosities $L_{\rm 2-10\,keV}>10^{44}\,\ergs$.

\section{Summary}
We studied the effect of dust on the Eddington limit in the molecular dusty torus of an AGN. While the Eddington limit for smooth dust distributions
is approximately 5 orders of magnitudes smaller than the classical Eddingtion limit for a fully ionized plasma, a clumpy dust torus provides a similar $\Ledd-\Mbh$-relation as the classical one, and is in good agreement with observed luminosities and black hole masses. The idea of a clumpy torus is based on self-gravitating,
optically thick dust clouds which are limited in size by the sheer of the gravitational potential of the central black hole. In the framework of this model,
we were able to derive a low-luminosity limit for the existence of an obscuring clumpy torus, which is of the order of $L\sim10^{42}\,\ergs$. Below this limit, the
physical and geometrical properties of the torus change significantly. Furthermore, we investigated the behaviour of the clumpy torus at high luminosities. We found that the largest clouds in the torus become gravitationally unbound to the central black hole if the AGN radiates close to the classical Eddington limit. In such a case, the dust clouds in the torus are no longer limited in size by the shear of the gravitational potential but by the AGN luminosity. The effective scale height of the radiation-limited tori decreases with luminosity, $H/R\propto L^{-1/4}$. The resulting $L$-dependence for the fraction of type 2 AGN, $f_2\propto L^{-0.25}$ is consistent with an analysis of several AGN surveys by \citet{Sim05}. We showed that the clumpy torus can account for broad-line AGN with high X-ray column densities, and that more such objects should be found at high rather than at low luminosities.

\section*{Acknowledgement}

We thank the anonymous referee for the much appreciated comments which helped to improve the paper significantly.

\bsp

\label{lastpage}

\end{document}